\documentclass[aps,preprint,amsmath,amssymb,showpacs,showkeys,nobibnotes]{revtex4}
\usepackage{mathrsfs}
\usepackage{amsfonts}
\usepackage{epsfig}
\usepackage{graphicx}
\def\wtilde#1{\setbox\z@\hbox{$\m@th#1$}%
 \ifdim\wd\z@>\tw@ em\mathaccent"0\msbfam@5D{#1}%
 \else\mathaccent"0365{#1}\fi}
\renewcommand{\thesection}{\arabic{section}}
\renewcommand{\theequation}{\thesection.\arabic{equation}}
\begin{document}

{\large \sf

\vspace*{2cm}
\title{
{\Huge \sf Deviations of the Lepton Mapping Matrix
from the Harrison-Perkins-Scott Form}
}
\author{
{\large \sf R. Friedberg$^1$ and T. D. Lee$^{1,2}$}\\
{\normalsize \it 1. Physics Department, Columbia University,}
{\normalsize \it New York, New York  10027, USA}\\
{\normalsize \it 2. China Center of Advanced Science and Technology
(CCAST/World Lab.),}
{\normalsize \it P. O. Box 8730, Beijing
100190, China}}
\vspace{1cm}

\vspace{1cm}

\begin{abstract}
\large \sf
We propose a simple set of hypotheses governing the
deviations of the leptonic mapping matrix from the
Harrison-Perkins-Scott (HPS) form. These deviations are supposed
to arise entirely from a perturbation of the mass
matrix in the charged lepton sector.
The perturbing matrix is assumed to be purely
imaginary (thus maximally $T$-violating) and to have a
strength in energy scale no greater (but perhaps smaller) than the
muon mass. As we shall show, it then follows
that the absolute value of the mapping matrix elements pertaining to the tau
lepton deviate by no more than $O((m_\mu/m_\tau)^2) \simeq 3.5 \times 10^{-3}$ from
their HPS values.

Assuming that  $(m_\mu/m_\tau)^2 $ can be neglected, we
derive two simple constraints on the four parameters
$\theta_{12}$, $\theta_{23}$, $\theta_{31}$, and $\delta$ of
the mapping matrix. These constraints are independent of the details
of the imaginary $T$-violating perturbation of the  charged lepton mass
matrix. We also show that the $e$ and $\mu$ parts of the mapping matrix have a
definite form governed by two parameters $\alpha$ and $\beta$;
any deviation of order $m_\mu/m_\tau $ can be accommodated by adjusting these two parameters.

\end{abstract}

\pacs{12.15.Ff, 11.30.Er, 14.60.-z}
\vspace{1cm}

\keywords{lepton mapping matrix; CP and T violation; Jarlskog invariant; timeon}

\maketitle

{\normalsize \sf

\large \sf
\newpage
\section*{\Large \sf 1.  Introduction}
\setcounter{section}{1} \setcounter{equation}{0} The last decade has
seen a well-defined situation take form with respect to neutrino
oscillations. The lepton mapping matrix is at least approximately
described by the "tribimaximal" formula of Harrison, Perkins and
Scott [1], and the differences of squared neutrino masses are known
to order of magnitude. The data on the mapping angles are so far
consistent with the HPS values, but best fits suggest some small
deviations. There is as yet no information on the $T$-violating
phase angle.

With respect to the mapping angles, the task of theoretical model
construction has been sorting itself into two directions: one
is to devise a natural way [2] in which the HPS formula can arise as a
zeroth approximation, and the other is to propose a perturbative
mechanism [3] that gives rise to deviations. This paper confines itself
to the second task.

In a recent paper [4], we suggested that $T$-violation in both
quarks and leptons could arise from the coupling of the Dirac matrix
$i\gamma_4\gamma_5$ with an undiscovered particle (called timeon) of
large mass. For leptons, it was proposed that the coupling occurs
only for the charged leptons, and without it the mapping matrix
would be exactly of the Harrison-Perkins-Scott [HPS] form. Both are
also assumed in this paper. As we shall see, many of the results of
the timeon paper can be derived without the additional assumptions
that the bare mass of the electron is zero and that the
$T$-violating coupling acts only on one vector in the flavor space.

The hypotheses proposed in this paper are thus a weaker subset
of those in [4]; these are

\noindent (i) The left-handed charged leptons are eigenstates of a hermitian matrix

\begin{equation}
L = L_0 + i L_1
\end{equation}
where $L_0$ and $L_1$ are real.

\noindent (ii) The "bare" charged leptons (i.e., eigenstates
of $L_0$) mix with neutrino precisely according to the Harrison-Perkins-Scott (HPS)
matrix.

\noindent (iii) the strength of $L_1$ is of order of the muon mass $m_\mu$ or less.

In Section 2, we shall show that assumptions (i)-(iii) lead to very small,
of the order of

\begin{equation}
({m_\mu \over m_\tau})^2 \cong 3.5 \times 10^{-3}
\end{equation}
deviations from HPS in the absolute values of three of the mapping matrix
elements

$$
|U_{31}|^2 = {1\over 6}~~, ~~~~~
|U_{32}|^2 = {1\over 3}
$$
and
\begin{equation}
|U_{33}|^2 = {1\over 2}
\end{equation}
Thus, there are two relations, to be discussed in Section 3, between three mapping angles
$\theta_{12}$, $\theta_{23}$, $\theta_{31}$ and the $T$ violating
phase $e^{i\delta}$ in the lepton mapping matrix.
These relations are valid to the accuracy of
order of $(m_\mu /m_\tau)$, but not to that of $(m_\mu /m_\tau)^2$.
Another consequence of (1.3) is that to the same accuracy, the
entire lepton mapping matrix can be described by two real
parameters, as will be summarized by the ($\alpha$, $\beta$) theorem
in Section 4. In Section 5, we shall discuss the experimental
implications of these relationships.

Throughout the paper the mapping angles
$\theta_{12},~\theta_{23},~\theta_{31}$ and the $T$-violating phase
$\delta$ are related to the mapping matrix elements $U_{ij}$ by
$$
\begin{array}{ll}
U_{11} = \cos \theta_{31} \cos \theta_{12},~~~~&
U_{12} = \cos \theta_{31} \sin \theta_{12}\\
U_{13} = \sin \theta_{31} e^{-i\delta},~~~~&
U_{23} = \sin \theta_{23} \cos \theta_{31}\\
\end{array}
$$
and
\begin{equation}
U_{33} = \cos \theta_{23} \cos \theta_{31}~.
\end{equation}

We shall write the physical charged lepton states as $|e>,~~|\mu>,~~|\tau>$
and the corresponding bare states as $|e_0>,~~|\mu_0>,~~|\tau_0>$.
The effect of the perturbation $iL_1$ is to cause the physical states
to differ from the corresponding bare states
by a unitary transformation $K=[K_{ll_0}]$ so that
\begin{equation}
|l> = \sum_{l_0} K_{ll_0} | l_0>
\end{equation}
with
\begin{equation}
K_{ll_0} = <l_0|l>
\end{equation}
where $l$, $l_0$ refer to $e,~\mu,~\tau $ and the corresponding
$e_0,~\mu_0,~\tau_0$. The free neutrino eigenstates will be called $|\nu_1>$,
$|\nu_2>$ and $|\nu_3>$ in the usual way. In the present proposal,
deviations from the HPS mapping matrix are due entirely to the perturbation
on the charged lepton mass matrix. Thus, the masses of the free neutrinos do not
affect these deviations. The neutrino masses do not play any role in this paper.

For convenience of notations, we shall introduce charged lepton stats $|1>$,
$|2>$, $|3>$ which are precisely related (without mixing) to the neutrino states
$|\nu_1>$, $|\nu_2>$, $|\nu_3>$ via the weak interaction. This enables us to write
for example $<1|e>$ for what is usually called $<\nu_1|\nu_e>$, and likewise
$<1|e_0>$ for $<\nu_1|\nu_{e_0}>$. The physical mapping matrix is then
\begin{equation}
U = [U_{lk}]
\end{equation}
where
\begin{equation}
U_{lk} = <k|l>
\end{equation}
with $k$ being $1,~2$, or $3$. It then follows from (1.5) that
\begin{equation}
U=KU_0
\end{equation}
or for the example of $k=1$ and $l=e$, the element $U_{e1}$ is
\begin{equation}
< 1 | e > =<1|e_0> <e_0|e> + <1|\mu_0><\mu_0|e> + <1|\tau_0><\tau_0|e>
\end{equation}
where the elements $<1|e_0>$, $<1|\mu_0>$ and $<1|\tau_0>$ refer to those
of $U_0$, and are precisely
the HPS matrix elements; i.e.,
\begin{equation}
\begin{array}{lll}
<1|e_0> = \sqrt{2\over 3},~~&
<2|e_0> = \sqrt{1\over 3},~~&
<3|e_0> = 0\\
<1|\mu_0> = -\sqrt{1\over 6},~~&
<2|\mu_0> = \sqrt{1\over 3},~~&
<3|\mu_0> = \sqrt{1\over 2}\\
<1|\tau_0> = \sqrt{1\over 6},~~&
<2|\tau_0> = -\sqrt{1\over 3},~~&
<3|\tau_0> = \sqrt{1\over 2}\\
\end{array}
\end{equation}

\newpage

\section*{\Large \sf 2. Effect of Large Tau Mass}
\setcounter{section}{2} \setcounter{equation}{0}
Consider the mapping element between the state
\begin{equation}
k=1,~2,~3
\end{equation}
and the $\tau$-state:
\begin{equation}
<k|\tau >=<k|e_0><e_0|\tau >+ <k|\mu_0 > <\mu_0|\tau >
+<k|\tau_0 ><\tau_0 |\tau >
\end{equation}
We shall compute $|<k|\tau >|^2$ to the accuracy
of $(m_\mu/ m_\tau)$, but neglecting corrections of order
$(m_\mu/ m_\tau)^2 $. By first-order perturbation
theory, we have
\begin{equation}
\begin{array}{ll}
<e_0|\tau > &\cong <e_0|i L_1|\tau_0> /(m_{\tau_0} - m_{e_0})\\
&\cong i<e_0|L_1|\tau_0>/m_\tau~~.
\end{array}
\end{equation}
Likewise,
\begin{equation}
<\mu_0|\tau >\cong i <\mu_0|L_1|\tau_0> /m_\tau~~.
\end{equation}
By hypothesis (iii), both these elements are of order of
$(m_\mu/m_\tau)$. Therefore
\begin{equation}
\begin{array}{ll}
1-|<\tau_0|\tau >|^2 &= |<e_0 |\tau >|^2 +|<\mu_0 |\tau>|^2\\
&\sim O[(m_\mu /m_\tau )^2 ] ~~.
\end{array}
\end{equation}
It follows that with neglect of $ O[(m_\mu /m_\tau )^2 ]$,
\begin{equation}
\begin{array}{ll}
<k|\tau > &= i <k |e_0 > <e_0|L_1|\tau_0>/ m_\tau  \\
& + i <k|\mu_0><\mu_0 |L_1 | \tau_0>/m_\tau + <k|\tau_0 >~~.
\end{array}
\end{equation}
By hypothesis (i), the elements of $L_1$ are real, and $<k|e_0>$,
$<k|\mu_0>$ and $<k|\tau_0> $ are also real since these are HPS
matrix elements. Thus, from (2.6) we have
\begin{equation}
\begin{array}{ll}
|<k|\tau >|^2 &= m_\tau^{-2} \left| <k|e_0 > <e_0|i L_1|\tau_0> + <k|\mu_0><\mu_0|i L_1|\tau_0>\right |^2\\
&~~~ +(<k|\tau_0>)^2\\
& = <k|\tau_0>^2 + O((m_\mu/m_\tau)^2)~~;
\end{array}
\end{equation}
i.e., with the neglect of $O(m_\mu/m_\tau)^2$,
$$
|<1|\tau>|^2 = {1\over 6}~~~, ~~~~~|<2|\tau>|^2 = {1\over 3}
$$
and
\begin{equation}
|<3|\tau>|^2 = {1\over 2}~~~,
\end{equation}
the same absolute values as HPS. (See also Eq. (12) of Xing [3].)

\newpage
\section*{\Large \sf 3. Consequences of the Model}
\setcounter{section}{3} \setcounter{equation}{0}

The standard form of the mapping matrix is
\begin{equation*}
\begin{array}{ll}
U &=
\left ( \begin{array}{ccc}
1 & 0 & 0\\
0 & c_{23} & s_{23}\\
0 & -s_{23} & c_{23}
\end{array} \right )
\left ( \begin{array}{ccc}
c_{31} & 0 & ~~s_{31}e^{-i\delta}\\
0 & 1 & 0\\
-s_{31}e^{i\delta} & 0 & c_{31}
\end{array} \right )
\left ( \begin{array}{ccc}
c_{12} & s_{12} & 0\\
-s_{12} & c_{12} & 0\\
0 & 0 & 1
\end{array} \right )\\
\end{array}
\end{equation*}
\begin{equation}
~~~~~=
\left ( \begin{array}{ccc}
c_{31}c_{12} & c_{31}s_{12} & s_{31}e^{-i\delta}\\
-s_{12}c_{23} - c_{12}s_{23}s_{31}e^{i\delta} &
c_{12}c_{23} - s_{12}s_{23}s_{31}e^{i\delta} &
s_{23}c_{31}\\
s_{12}s_{23} - c_{12}c_{23}s_{31}e^{i\delta} &
-c_{12}s_{23} - s_{12}c_{23}s_{31}e^{i\delta} &
c_{23}c_{31}\\
\end{array} \right )
\end{equation}
with
\begin{equation}
s_{ij} = \sin\theta_{ij}  ~~~~\text{and}~~~c_{ij} = \cos \theta_{ij}~~~.
\end{equation}
Eq. (2.8) can then be written as
\begin{equation}
|s_{12}s_{23} - c_{12}c_{23}s_{31}e^{i\delta}|^2 = |U_{31}|^2 = {1\over 6}~~,
\end{equation}
\begin{equation}
|-c_{12}s_{23} - s_{12}c_{23}s_{31}e^{i\delta}|^2 = |U_{32}|^2 = {1\over 3}
\end{equation}
and
\begin{equation}
c_{23}^2 c_{31}^2  = |U_{33}|^2 = {1\over 2}~~.
\end{equation}
(Here, $U_{3j}$ is the same $U_{\tau j} = <j|\tau>$ of
previous sections, and likewise for other $U_{ij}$.)

It is convenient to express relations in terms of quantities
that vanish in the HPS limit. From (3.5), we find
\begin{equation}
c_{23}^2 s_{31}^2  = {1\over 2} \tan^2\theta_{31}
\end{equation}
and
\begin{equation}
(2 c_{23}^2 - 1) c_{31}^2 = 1 - c_{31}^2 = s_{31}^2~~~,
\end{equation}
which on division by $c_{31}^2$ gives
\begin{equation}
\cos 2 \theta_{23} = \tan^2 \theta_{31}~~.
\end{equation}
Note that both sides of (3.6)-(3.8) vanish at the HPS point.

Next, the difference of (3.3) and (3.4) gives
$$
\begin{array}{ll}
\displaystyle {1\over 6}  &= |U_{32}|^2 - |U_{31}|^2\\
&= |-c_{12}s_{23} - s_{12}c_{23}s_{31}e^{i\delta}|^2 -
|s_{12}s_{23} - c_{12}c_{23}s_{31}e^{i\delta}|^2\\
\end{array}
$$
\begin{equation}
=( s_{23}^2 - c_{23}^2 s_{31}^2)(c_{12}^2 - s_{12}^2) +
4c_{12}s_{12}c_{23}s_{23}s_{31}\cos \delta
\end{equation}
From (3.5), we have
\begin{equation}
\begin{array}{ll}
s_{23}^2 - c_{23}^2 s_{31}^2
&= s_{23}^2 - c_{23}^2 (1-c_{31}^2)\\
&= {1\over 2} - \cos 2 \theta_{23}~~.
\end{array}
\end{equation}
and
\begin{equation}
(c_{23}s_{23}s_{31})^2 =
(c_{23}s_{23}c_{31}\tan\theta_{31})^2 ={1\over 2}
(s_{23}\tan\theta_{31})^2 ~~,
\end{equation}
which on account of (3.8) can also be written as
\begin{equation}
(c_{23}s_{23}s_{31})^2 =
{1\over 2} s_{23}^2 \cos 2\theta_{23} =
{1\over 4} (1 - \cos 2\theta_{23}) \cos 2\theta_{23}
\end{equation}
Using (3.10)-(3.12), we may write (3.9) as an equation of
$\theta_{12}$, $\theta_{23}$  and $\delta$,
\begin{equation}
{1\over 6} =
({1\over 2} - \cos 2 \theta_{23}) \cos 2 \theta_{12} +
[ (1-\cos 2 \theta_{23})\cos 2 \theta_{23} ]^{1\over 2} \sin 2 \theta_{12}
\cos \delta
\end{equation}

To obtain a relation free of square roots, we may shift the term containing
$\cos 2\theta_{12}$ to the left-hand side, multiply the equation by 2 and then square both sides.
This yields
\begin{equation}
[{1\over 3}- (1-2 \cos 2 \theta_{23})\cos 2 \theta_{12} ]^2
=4 (1-\cos 2 \theta_{23})\cos 2 \theta_{23}\sin^2 2\theta_{12}\cos^2 \delta ~~,
\end{equation}
which can be solved as a quadratic equation in either
$\cos 2\theta_{12}$ or $\cos 2\theta_{23}$, supposing that the other is given
as well as $\cos^2\delta$. Eqs (3.8) and (3.14) may be taken as two useful equations relating
$\theta_{23}$ to $\theta_{31}$, as well as
$\theta_{12}$ to $\theta_{23}$ and $\delta$.
Both relations follow from hypotheses (i)-(iii) stated in
Section 1, and are accurate to the accuracy of $(m_\mu/m_\tau)$.

For certain purposes, a further simplification can be achieved.
Define $\theta_{12}^{HPS}$
to be the HPS value of $\theta_{12}$, so that
\begin{equation}
\cos 2 \theta_{12}^{HPS} = {1\over 3}~~.
\end{equation}
Introduce a positive angle $\phi$ such that
\begin{equation}
\sin^2 \phi = \cos 2 \theta_{23}~~.
\end{equation}
Then the square root of (3.12) can be written as
\begin{equation}
s_{23}c_{23}s_{31} = {1\over 2 } \sin \phi \cos \phi
\end{equation}
and (3.13) gives, after  being multiplied by 2,
\begin{equation}
{1\over 3} = \cos 2 \phi \cos 2\theta_{12} +\sin 2 \phi
\sin 2 \theta_{12} \cos \delta~~.
\end{equation}

On account of (3.15), we may write this as
\begin{equation}
\cos 2 \theta_{12}^{HPS} = \cos 2 \phi \cos 2 \theta_{12} +
\sin 2\phi \sin 2\theta_{12} \cos \delta ~~,
\end{equation}
which is precisely the law of cosines for a spherical triangle,
as shown in Figure 1.
The sides of the triangle are
\begin{equation}
2\theta_{12}, ~~~2\phi ~~~\text{and}~~~2\theta_{12}^{HPS}~,
\end{equation}
and $\delta$ is the angle between $2\theta_{12}$ and $2\phi$.
(Note that $\theta_{23}$ and $\theta_{31}$ are explicit functions
of $\phi$ through (3.16) and (3.8)).

In the absence of $T$ violation, we have
\begin{equation}
\cos \delta =\pm 1
\end{equation}
and correspondingly,
\begin{equation}
\theta_{12} = \theta_{12}^{HPS} \pm \phi~~.
\end{equation}
In the presence of $T$ violation, we may write (3.19) as
\begin{equation}
\begin{array}{l}
\left[
{1\over 2} ( 1 + \cos \delta) +
{1\over 2} ( 1 - \cos \delta) \right ] \cos 2 \theta_{12}^{HPS}\\
~~~~~~= {1\over 2} ( 1 + \cos \delta)\cos 2 (\theta_{12}-\phi )
+{1\over 2} ( 1 - \cos \delta)\cos 2 (\theta_{12}+\phi )
\end{array}
\end{equation}
which, in turn, leads to
\begin{equation}
{1 + \cos\delta \over 1 - \cos \delta} =
{\cos 2 \theta_{12}^{HPS} - \cos 2 (\theta_{12} + \phi) \over \cos 2 (\theta_{12} - \phi)
-\cos 2 \theta_{12}^{HPS} }
\end{equation}
and, on account of (3.15)
\begin{equation}
{1 + \cos\delta \over 1 - \cos \delta} =
{1 - 3 \cos 2 (\theta_{12} + \phi) \over 3 \cos 2 (\theta_{12} - \phi)
- 1 } ~~~.
\end{equation}
The above LHS is an increasing function of $\cos \delta$, and its RHS at
fixed $\phi$ is an increasing function of $\theta_{12}$.
Thus,
\begin{equation}
\left ( {\partial \theta_{12} \over \partial \cos \delta } \right )_{\theta_{23}} =
\left ( {\partial \theta_{12} \over \partial \cos \delta } \right )_{\theta_{31}} =
\left ( {\partial \theta_{12} \over \partial \cos \delta } \right )_\phi > 0 ~~~.
\end{equation}
From (3.8), (3.16), (3.22), (3.26) and by eliminating $\delta$,
we obtain the following statement relating the three mapping angles.
\begin{equation}
\cos 2 \theta_{23} = \tan^2 \theta_{31} =\sin^2 \phi \geq \sin^2 (\theta_{12} - \theta_{12}^{HPS} )
\end{equation}
where in the last relation, the inequality holds for $\cos^2 \delta < 1$, and the equality when $\cos^2 \delta = 1$.

The Jarlskog invariant $J$ [5] is given by
\begin{equation}
J = s_{12} c_{12} s_{23} c_{23} s_{31} c_{31}^2 \sin \delta~~.
\end{equation}
From (3.17), we can also write
\begin{equation}
J = {1\over 8} \sin 2\theta_{12} \sin 2 \phi \cos^2 \theta_{31} \sin \delta~~~.
\end{equation}

\newpage
\section*{\Large \sf 4.  The alpha-beta Theorem}
\setcounter{section}{4} \setcounter{equation}{0}
This section is devoted to establishing a theorem that shall be called the alpha-beta
theorem.

\noindent \underline{Theorem}: Suppose the mapping matrix $U$ to have its $\tau$-elements
given (apart from their phases) by the HPS values
$$
|U_{\tau 1}|^2 = {1\over 6}~~, ~~~~~|U_{\tau 2}|^2 = {1\over 3}~,
$$
\begin{equation}
|U_{\tau 3}|^2 = {1\over 2}
\end{equation}
as in (3.3)-(3.5), and that the third $\tau$-element is real and positive with
\begin{equation}
U_{\tau 3} = {1\over\sqrt{ 2}}~~~.
\end{equation}
Then there exist real numbers $\alpha$ and $\beta$, such that
\begin{equation}
U=S_1^{-1} V S_2
\end{equation}
where $S_1$ and $S_2$ are both diagonal unitary matrices, and
\begin{equation}
V = \left( \begin{array}{ccc}
\sqrt{2\over 3}\sin {\alpha\over 2} +\sqrt{1\over 6}\cos {\alpha\over 2}e^{i\beta} ~~&
\sqrt{1\over 3}\sin {\alpha\over 2} -\sqrt{1\over 3}\cos {\alpha\over 2}e^{i\beta}~~&
-\sqrt{1\over 2}\cos {\alpha\over 2} e^{i\beta}\\[0.2cm]
-\sqrt{1\over 6}\sin {\alpha\over 2} +\sqrt{2\over 3}\cos {\alpha\over 2}e^{-i\beta} ~~&
\sqrt{1\over 3}\sin {\alpha\over 2} + \sqrt{1\over 3}\cos {\alpha\over 2}e^{-i\beta}~~&
\sqrt{1\over 2}\sin {\alpha\over 2} \\[0.2cm]
\sqrt{1\over 6}& -\sqrt{1\over 3}&\sqrt{1\over 2}
\end{array}\right)
\end{equation}

To prove the theorem, we make use of the following lemma, proved in Appendix B.

\noindent \underline{Lemma}: Let $W$ be a $3\times 3$ unitary matrix
of the form
\begin{equation}
W=\left ( \begin{array}{rr}
t &~~~ \xi  \\[0.2cm]
\tilde{\eta} & d
\end{array}
\right )
\end{equation}
where $t$ is a $2\times 2$ matrix,  $\xi$ and
$\eta$ are both real $2\times 1$ column matrices  and $d$ a real number.
Then $t$ can be written in terms of $\xi$,
$\eta$, $d$ and an extra real parameter $\beta$ by the formula
\begin{equation}
t = (1 - d^2)^{-1} (-d \xi \tilde{\eta} + \xi'\tilde{\eta'} e^{-i\beta})
\end{equation}
where $\xi'$ and $\eta'$ are both real $2\times 1$ column matrices satisfying
$$
\tilde{\xi'} \xi' = \tilde{\xi}\xi~,~~~~
\tilde{\eta'}\eta' = \tilde{\eta}\eta
$$
and
\begin{equation}
\tilde{\xi'} \xi = \tilde{\eta'}\eta = 0~~.
\end{equation}

Supposing the Lemma to be established, we prove
the alpha-beta theorem  as follows:

The five matrix elements in the third row
and the third column of $U$ can all be made real by introducing
an extra phase factor into each of these elements. This task can
be achieved by introducing unitary diagonal matrices $S'_1$
and $S'_2$ such that
\begin{equation}
W = S'_1 U {S'_2}^{-1}
\end{equation}
has the form $(4.5)$ required by the lemma.
Moreover, for our applications,
\begin{equation}
\eta=\left ( \begin{array}{c}
\sqrt {1\over 6} \\
-\sqrt {1\over 3} \\
\end{array}\right )
\end{equation}
and
\begin{equation}
d = \sqrt {1\over 2}~~~.
\end{equation}
The corresponding vector $\eta'$ is, in accordance with (4.7),
\begin{equation}
\eta' =\left ( \begin{array}{c}
-\sqrt {1\over 3} \\
-\sqrt {1\over 6} \\
\end{array}\right )
\end{equation}
with the signs in $\eta$ and $\eta'$ being chosen for later
convenience; the ambiguity will be subsumed in the arbitrariness of $\beta$
in (4.6).

Since $W$ is unitary, we have
\begin{equation}
t^\dagger t +\eta \tilde{\eta} = 1
\end{equation}
\begin{equation}
\tilde{\xi} t + d \tilde{\eta} = 0
\end{equation}
and
\begin{equation}
\tilde\xi \xi = 1- d^2 = {1\over 2}~~.
\end{equation}
Hence, we may define
\begin{equation}
\xi =\sqrt{1\over 2} \left ( \begin{array}{c}
\cos {\alpha \over 2} \\
\sin {\alpha \over 2} \\
\end{array}\right )
\end{equation}
and on account of (4.7)
\begin{equation}
\xi' =\sqrt{1\over 2} \left ( \begin{array}{c}
\sin {\alpha \over 2} \\
-\cos {\alpha \over 2} \\
\end{array}\right )
\end{equation}
Substituting these expressions into (4.5)-(4.6), we find
\begin{equation}
t = - \left ( \begin{array}{cc}
\sqrt{1\over 6}\cos {\alpha\over 2}+ \sqrt{2\over 3}\sin {\alpha\over 2} e^{-i\beta} &
-\sqrt{1\over 3}\cos {\alpha\over 2}+ \sqrt{1\over 3}\sin {\alpha\over 2} e^{-i\beta} \\
-\sqrt{1\over 6}\cos {\alpha\over 2}+ \sqrt{2\over 3}\cos {\alpha\over 2} e^{-i\beta} &
\sqrt{1\over 3}\sin {\alpha\over 2}+ \sqrt{1\over 3}\cos {\alpha\over 2} e^{-i\beta} \\
\end{array}\right )
\end{equation}
Assembling  $W$ according to (4.5), we find that the matrix
\begin{equation}
V \equiv  \left ( \begin{array}{ccc}
-e^{i\beta} & 0 & 0\\
0 & 1 & 0\\
0 & 0 & 1\\
\end{array}\right ) W = S_1^{''}W
\end{equation}
is given by (4.4), and that establishes the alpha-beta theorem, with
\begin{equation}
S_1 = S_1^{''} S_1^{'}
\end{equation}
and
\begin{equation}
S_2 = S_2^{'}~~.
\end{equation}

\newpage
By using the alpha-beta theorem, we can derive several interesting relations
between the four parameters $\theta_{12}, ~~\theta_{23}, ~~\theta_{31}$
and $\delta$ of the mapping matrix $U$. These will be discussed in Appendix $C$.

\noindent\underline{Remark}
It will be seen that the above expression for $V$ is identical to the matrix
$V_{l-map}$ shown  in Table 1 of [4], when correction terms in the quantities
\begin{equation}
\chi_e,~~\chi_p, ~~c_e,~~c_p
\end{equation}
are neglected; the $"\chi_e,~~\chi_p"$ quantities will be shown in Appendix A
of this paper to leave the absolute values of the matrix elements in (4.1) unchanged except
by an amount of $O((m_\mu/m_\tau)^2)$. It follows therefore, from the alpha-beta
theorem just established, that the $"c_e, ~c_p"$ correction terms in the upper two
rows of $V_{l-map}$ in [4], which are admittedly of first order in $(m_\mu/m_\tau)$,
can be taken into account (to that order) by adjusting the values of $\alpha$ and $\beta$,
which in Ref. [4] were restricted to be certain given expressions in terms of
the detailed matrices $G$ and $F$.

The outcome is that any experimental predictions made from using Table 1
of [4], plus the knowledge that its $"c_e,~c_p"$-corrections are of first
order and its $\chi$-corrections of second order in $m_\mu/m_\tau$,
can just as well be made on the basis of the weaker hypotheses (i)-(iii)
stated in Section 1 of this paper.

\newpage
\section*{\Large \sf 5.  Discussion}
\setcounter{section}{5} \setcounter{equation}{0}
(i) In the HPS limit, from (1.11) and (3.1)
$$
\sin \theta_{12}^{HPS} = \sqrt{ 1\over 3}~~~~,~~~
\sin \theta_{31}^{HPS} = 0
$$
and
\begin{equation}
\sin \theta_{23}^{HPS} = \sqrt{ 1\over 2}~~.
\end{equation}
From (3.8), we have
\begin{equation}
1- \sin^2 2 \theta_{23} = \tan^4 \theta_{31}~~~.
\end{equation}
A striking feature of our model is that it predicts a much smaller
deviation from HPS in $\theta_{23}$ than in $\theta_{31}$.
Since $\theta_{31}$ is known to be small, from (5.2) we expect $\theta_{23}$ even closer to its
HPS value of $45^\circ$, as a linear deviation in $\theta_{23}$ would be quadratic in $\theta_{31}$.

At present, current data [6-10] are compatible (within $1\sigma$)
with the HPS values of $\theta_{23}$ and $\theta_{31}$, but there is
a suggestion that $\sin^2 \theta_{31}$ may be about $0.015$. If we
take this value, then
\begin{equation}
\sin^2 2 \theta_{31} = 0.0591
\end{equation}
and from (5.2)
\begin{equation}
\sin^2 2 \theta_{23} = 0.9998~~.
\end{equation}
These data seem not yet precise enough to say whether the deviation
of $\sin^2 2\theta_{23}$ from $1$ is as small as given by (5.4).

(ii) Next, we turn to our second relation, (3.16) and (3.18) relating $\theta_{12}$
to $\theta_{23}$ and $\delta$. We may replace $\cos 2\theta_{23}$ with
$\tan^2 \theta_{31}$ in accordance with (3.8).
At any fixed $\delta$, these equations define a curve describing the variations of
\begin{equation}
x =\sin^2 \theta_{12} ~~~\text{vs}~~~
y =\sin^2 \theta_{31} ~~~.
\end{equation}
The envelope of the family of such curves is shown in Figure 2, and corresponds to
\begin{equation}
\cos \delta = \pm 1~~~.
\end{equation}
The region below the envelope corresponds to $\cos^2 \delta > 1$ and is therefore forbidden.

An examination of current data [10-13]
indicates that points on the outermost curve (no $T$ violation) are
far from the best fit, and that the forbidden
region below the curve is improbable.
As the best fit (represented by the circle) shown in Figure 2 already prefers large $T$ violation, a measurement
of $\delta$, combined with improved precisions in $\theta_{12}$ and $\theta_{31}$, would
give a sensitive test to our model.

(iii) It is of interest to compare the assumptions and results of Ge, He and Yin
[GHY, ref.14] and those of this paper.
Both papers regard the HPS mapping matrix as correct to 0$^{th}$ order, and
concentrate on the 1st-order deviations from it. In GHY, these deviations are attributed
to a perturbation in the neutrino sector, whereas in the present paper the perturbation arises in
the charged lepton sector.

In the notations of this paper, a perturbation in the charged lepton sector leads
to a mapping matrix $U$ given by (1.9)
\begin{equation*}
U=KU_0~~,
\end{equation*}
whereas a perturbation in the neutrino sector would yield an equivalent form
\begin{equation}
U=U_0 K' ~~.
\end{equation}
A difference  appears only when different physical approximations are made in $K$ and $K'$.
As a result, the constraints arrived at on the four parameters $\theta_{12}$, $\theta_{23}$,
$\theta_{31}$ and $\delta$ can be quite different. Their result [GHY(4.87a)] in our notation is
\begin{equation}
\theta_{23} - 45^\circ \cong -\theta_{31} \cot \theta_{12} \cos \delta
\end{equation}
or (to leading order in deviation from HPS)
\begin{equation}
\cos 2 \theta_{23} \cong 2\sqrt{2} \tan \theta_{31} \cos\delta~~~.
\end{equation}
This differs from our (3.8) in two important ways: our model
gives a relation between $\theta_{23}$ and $\theta_{31}$ independent of $\delta$,
and it makes $\cos 2 \theta_{23}$ quadratic in $\tan \theta_{31}$ instead
of linear. Thus, unless $\tan \theta_{31} \cong 2\sqrt{2}\cos \delta$,
experiments now under way [15-18] could lead to a resolution between the
hypothesis of charged lepton perturbation (this paper) and that of neutrino-perturbation (GHY).

\newpage
\section*{\Large \sf Appendix A}
\renewcommand{\thesection}{\Alph{section}}
\renewcommand{\theequation}{\thesection.\arabic{equation}}
\setcounter{section}{1}
\setcounter{equation}{0}

In the timeon model [4], the three left-handed physical leptons are eigenvectors
of a hermitian matrix
\begin{equation}
(G+iF)(G-iF) = G^2 + i[F,G]+F^2
\end{equation}
where $G$ and $F$ are both real and symmetric. The eigenvalues of $(G+iF)(G-iF)$ are
the squares of the physical masses, with the corresponding bare charged leptons
the eigenvectors of $G$ (or $G^2$). This is a slightly more involved situation than
the one described in the present paper, but it leads to the same results.

Set
$$
L_0 = G^2~~~~,~~~~ L_1 = i[F, G]
$$
and
\begin{equation}
L_2 = F^2~~~.
\end{equation}
The bare states are eigenvectors of $L_0$, and the corresponding physical
ones, those of
\begin{equation}
L = L_0 + L_1 + L_2~~~.
\end{equation}
The perturbation caused by $L_2$ on the state $|\tau>$ is of the order
\begin{equation}
\left ( {F\over m_\tau }\right )^2
\end{equation}
and therefore negligible. The perturbation caused by $L_1$ is of the
order
\begin{equation}
{ F\over m_\tau }
\end{equation}
and purely imaginary. Thus, by following
the discussion given in Section 2, we can readily arrive at the
inference that the deviation of $|<k | \tau>|$ from
$|<k | \tau_0>|$ is of order $( {F/ m_\tau })^2$.

\newpage
\section*{\Large \sf Appendix B}
\renewcommand{\thesection}{\Alph{section}}
\renewcommand{\theequation}{\thesection.\arabic{equation}}
\setcounter{section}{2}
\setcounter{equation}{0}
Here we prove the lemma  stated in Sec. 4.
Let $W$ be the unitary matrix given by (4.5);
it follows then
\begin{equation}
1 = W^\dagger W =
\left (\begin{array}{ll}
t^\dagger t + \eta\tilde{\eta}~~~~&
t^\dagger \xi + d{\eta}\\
\tilde\xi t + d \tilde{\eta}&
\tilde{\xi}\xi + d^2
\end{array}\right )
\end{equation}
and
\begin{equation}
1 = W W^\dagger =
\left (\begin{array}{ll}
t t^\dagger  + \xi\tilde{\xi}~~~~&
t \eta + d{\xi}\\
\tilde{\eta} t^\dagger + d \tilde{\xi}&
\tilde{\eta}\eta + d^2
\end{array}\right )
\end{equation}
in which $t$ is a $2\times 2$ matrix, $\xi$ and $\eta$
are both real $2\times 1$ column matrices and $d$ a real number.
From the above equations, we have from the lower diagonal elements
\begin{equation}
\tilde{\xi}\xi = \tilde{\eta}\eta = 1 - d^2
\end{equation}
and from the off-diagonal elements
\begin{equation}
\begin{array}{l}
\tilde{\xi}t + d \tilde{\eta} = 0~~~,\\
t \eta + d \xi = 0~~~.
\end{array}
\end{equation}
Let $\xi'$ and $\eta'$ be the two real column matrices that satisfy (4.7).
We observe that the four products
\begin{equation}
\xi\tilde{\eta}~,~~~\xi\tilde{\eta'}~,~~~\xi'\tilde{\eta}~~~\text{and}~~~\xi'\tilde{\eta'}
\end{equation}
form a complete basis for $2\times 2$ matrices. Thus, we can express
\begin{equation}
t = t_{11} \xi \tilde{\eta} + t_{12} \xi \tilde{\eta'}
+ t_{21} \xi' \tilde{\eta}+ t_{22}\xi' \tilde{\eta'}
\end{equation}
in which $t_{11}$, $\cdots$, $t_{22}$ are four
coefficients.

Combining (B.6) with (B.3) and (4.7), we have
\begin{equation*}
\tilde{\xi} t = t_{11}(\tilde{\xi}\xi )\tilde{\eta} + t_{12} (\tilde{\xi}\xi)\tilde{\eta'}
\end{equation*}
and
\begin{equation}
 t \eta = t_{11}\xi (\tilde{\eta}\eta ) + t_{21} \xi'(\tilde{\eta}\eta)~~.
\end{equation}
Combining these two equations with (B.4), we find
$$
t_{12} = t_{21} = 0
$$
and
\begin{equation}
t_{11} = - { d\over 1 - d^2}
\end{equation}
It will be convenient to write the coefficient $t_{22}$ as
\begin{equation}
t_{22} = \lambda t_{11}
\end{equation}
with $\lambda$ an unknown complex number. Thus, we can write
\begin{equation}
t =  - { d\over 1 - d^2}(\xi \tilde{\eta} + \lambda \xi' \tilde{\eta'})~~.
\end{equation}

Turn now to the upper left part of
$W^\dagger W$; it gives
\begin{equation}
t^\dagger t  + \eta \tilde{\eta} = I
\end{equation}
where $I$ is the $2\times 2$ unit matrix. From (B.10), we find
\begin{equation}
t^\dagger t =  ({ d\over 1 - d^2})^2 (1 - d^2) ( \eta\tilde{\eta} + |\lambda|^2 \eta' \tilde{\eta'})
\end{equation}
where we have used (B.3) and (4.7) to eliminate the inner products in $\xi$ and $\xi'$.
Using (B.11) and (B.12) and after some rearrangement, we have
\begin{equation}
\eta\tilde{\eta} + d^2 |\lambda|^2 \eta' \tilde{\eta'} = (1-d^2)I~~.
\end{equation}
On the other hand, we can also verify that
\begin{equation}
\eta\tilde{\eta} + \eta'\tilde{\eta'} = (1-d^2)I~~.
\end{equation}
by multiplying both sides on the right alternatively by $\eta$ and by
$\eta'$.

Thus, (B.13) and (B.14) lead to
\begin{equation}
d^2|\lambda|^2 = 1~~.
\end{equation}
This enables us to introduce a phase factor
\begin{equation}
e^{-i\beta} = -\lambda d~~~,
\end{equation}
so that (B.10) becomes (4.6), and the lemma is established.

\newpage
\section*{\Large \sf Appendix C}
\renewcommand{\thesection}{\Alph{section}}
\renewcommand{\theequation}{\thesection.\arabic{equation}}
\setcounter{section}{3}
\setcounter{equation}{0}
In this Appendix, we derive certain relations between the angles
$\theta_{12}$, $\theta_{23}$, $\theta_{31}$, and $\delta$ by a
different route, making use of the alpha-beta theorem.

Let
\begin{equation}
U = U(\theta_{12}, ~\theta_{23}, ~\theta_{31}, ~\delta)~
\end{equation}
and
\begin{equation}
V = V(\alpha,~\beta)
\end{equation}
be the matrices given by (3.1) and (4.4), and $J_U$ and $J_V$,
their respective Jarlskog invariants. From (3.28) and (4.4), we
find
\begin{equation}
J_U = s_{12}c_{12}s_{23}c_{23} s_{31}c_{31}^2 \sin\delta ~~~,
\end{equation}
\begin{equation}
J_V = {1 \over 12} \sin\alpha \sin\beta~~~.
\end{equation}
Here, we assume (4.1) and therefore (by the alpha-beta
theorem) also (4.3). It follows then
\begin{equation}
J_U = J_V
\end{equation}
and for all $(i,j)$,
\begin{equation}
|U_{ij}| = |V_{ij}|~~.
\end{equation}
Denote
$$
\begin{array}{lll}
a = \theta_{23}~~, ~~~~~~~& b = \theta_{31}~~,~~~~~~~&c=\theta_{12}\\[0.2cm]
s_a = \sin \theta_{23}~~, & s_b = \sin \theta_{31}~~, & s_c = \sin \theta_{12}~~,\\[0.2cm]
c_a= \cos \theta_{23}~~, & c_b = \cos \theta_{31}~~, & c_c = \cos \theta_{12}~~,\\[0.4cm]
\end{array}
$$
\begin{equation}
\Gamma = \cos 2 \theta_{12} = c_c^2 - s_c^2
\end{equation}
and therefore
\begin{equation}
1-\Gamma^2 = \sin^2 2\theta_{12} = 4 s_c^2 c_c^2~~~.
\end{equation}
We shall explore the consequences of eliminating successively $\alpha$, $\delta$
and $\beta$ from (C.5)-(C.6).

\noindent (i) Determinations of $\cos \delta$ and $\cos \beta$

By equating
\begin{equation}
|U_{32}|^2 - |U_{31}|^2 = |V_{32}|^2 - |V_{31}|^2 = {1 \over 6}~~~,
\end{equation}
we find $\cos\delta$ given by
\begin{equation}
[1-6 \Gamma(s_a^2 - c_a^2 s_b^2)]^2 = (1-\Gamma^2)(12s_a c_a s_b \cos \delta)^2~~~.
\end{equation}
Likewise, from
\begin{equation}
|U_{13}|^2 = |V_{13}|^2 ~~~~~\text{and}~~~~|U_{23}|^2 = |V_{23}|^2~~~~,
\end{equation}
it follows then
\begin{equation}
\sin^2 \alpha = (4 s_a s_b c_b)^2~~~~,
\end{equation}
and from $|U_{12}|^2 = |V_{12}|^2$, we find
\begin{equation}
\sin^2 \alpha \cos^2 \beta = (1 - 3 c_b^2 s_c^2)^2~~.
\end{equation}
Thus,
\begin{equation}
\cos^2 \beta = [(1 - 3 c_b^2 s_c^2 ) /4 s_a s_b c_b]^2~~.
\end{equation}

\noindent (ii) Relation between $\sin \beta$ and $\sin \delta$

From $|U_{33}|^2 = |V_{33}|^2$,
\begin{equation}
c_a^2 c_b^2 = {1\over 2}
\end{equation}
which together with the equality of Jarlskog invariants (C.3) and (C.4) yield
\begin{equation}
2 \sin^2 \beta = 9 s_c^2 c_c^2 \sin^2 \delta~~.
\end{equation}

\noindent (iii) Elimination of $\delta$

Multiplying (C.16) by $64 s_a^2 c_a^2 s_b^2$, we have
\begin{equation}
\begin{array}{ll}
128 s_a^2 c_a^2 s_b^2\sin^2\beta &= 144 (4 s_c^2 c_c^2) s_a^2 c_a^2 s_b^2 \sin^2 \delta\\
&= 144 (1 - \Gamma^2 ) s_a^2 c_a^2 s_b^2 \sin^2 \delta
\end{array}
\end{equation}
with $\Gamma$ given by (C.7). Combining the above equation with (C.10), we derive
\begin{equation}
[1-6\Gamma (s_a^2 - c_a^2 s_b^2)]^2 +128 s_a^2 c_a^2 s_b^2 \sin^2\beta = 144 (1 - \Gamma^2 ) s_a^2 c_a^2 s_b^2~~~.
\end{equation}

\noindent (iv) Elimination of $\beta$

Multiplying (C.14) by $128 s_a^2 c_a^2 s_b^2$, we find
\begin{equation}
\begin{array}{ll}
128 s_a^2 c_a^2 s_b^2 \cos^2\beta &= 8 (1-3 c_b^2 s_c^2) (c_a/c_b)^2\\
&= 16 (1-3 c_b^2 s_c^2) c_a^4
\end{array}
\end{equation}
on account of (C.15). The sum of (C.18) and (C.19) gives
\begin{equation}
[1-6\Gamma (s_a^2 - c_a^2 s_b^2)]^2 +128 s_a^2 c_a^2 s_b^2 = 144 (1 - \Gamma^2 ) s_a^2 c_a^2 s_b^2 + (4c_a^2 - 6 s_c^2)^2
\end{equation}
and therefore a relation between the angles $a,~b$ and $c$:
\begin{equation}
[1-6\Gamma (s_a^2 - c_a^2 s_b^2)]^2 - (4c_a^2 - 6 s_c^2)^2 =  16 (1 - 9 \Gamma^2 ) s_a^2 c_a^2 s_b^2~~~.
\end{equation}

It may appear that by combining (C.21) with (3.8) one could arrive
at a determination of $a$ and $b$  in terms of $c$ (i.e., of
$\theta_{23}$ and $\theta_{31}$ in terms of $\theta_{12}$), without
fixing $\delta$.
But as we shall show, (C.21) and (3.8) are actually redundant.

Define
\begin{equation}
X \equiv 1-6\Gamma (s_a^2 - c_a^2 s_b^2)~~~,
\end{equation}
\begin{equation}
Y \equiv 4c_a^2 - 6 s_c^2
\end{equation}
and
\begin{equation}
Z \equiv 16 (1 - 9 \Gamma^2 ) s_a^2 c_a^2 s_b^2~~~.
\end{equation}
with $\Gamma = \cos 2 c$ given by (C.7). Thus (C.21) becomes
\begin{equation}
X^2 - Y^2 = Z~~~.
\end{equation}

From (3.8), we have $\cos 2 a = \tan^2 b$ and therefore
\begin{equation}
2c_a^2 c_b^2 = 1~~,
\end{equation}
Hence, we can express $b$ in terms of $a$, $c$ in terms of $\Gamma$
and write (C.22)-(C.24)
as
\begin{equation}
X + Y = 2 ( 2c_a^2 - 1)(1 + 3\Gamma)~~~,
\end{equation}
\begin{equation}
X - Y = -4  ( c_a^2 - 1)(1 - 3\Gamma)~~~,
\end{equation}
and
\begin{equation}
Z = -8 ( 2c_a^2 - 1)( c_a^2 - 1)(1 - 9\Gamma^2)~~~.
\end{equation}
It follows then (C.25) is an identity.

\newpage

\begin{figure}[h]
\centerline{
\epsfig{file=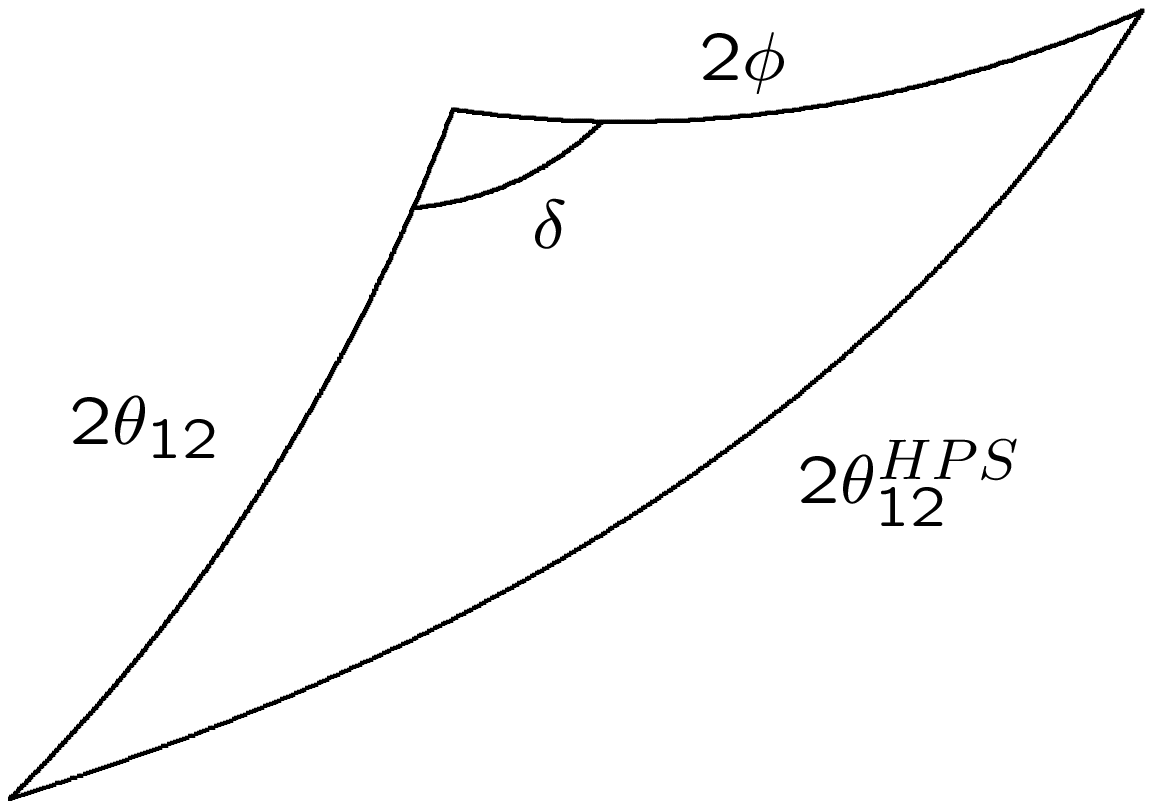, width=11cm, height=6.8cm}}
\vspace{.5cm}
 \centerline{\large \sf The spherical triangle described by (3.19),
with $\sin^2 \phi = \cos 2 \theta_{23} = \tan^2\theta_{31}$. }
\vspace{.5cm}
\centerline{\Large Figure 1}
\end{figure}

\newpage
\begin{figure}
\centerline{ \epsfig{file=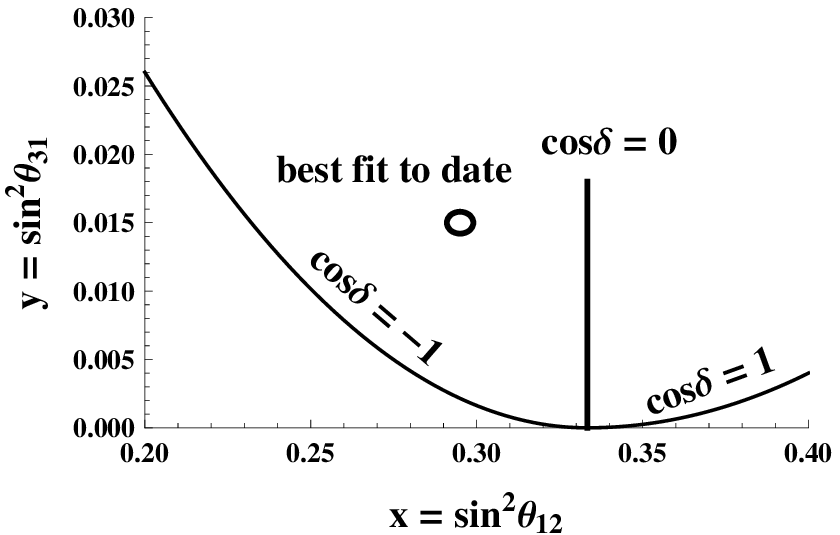, width=16cm,
height=10.0cm}} \vspace{.5cm} \centerline{{\large\sf The curve
described by (5.5) and (5.6).}} \centerline{{\large\sf Points below
the curve are forbidden and the HPS limit is $(x,y) = ({1\over 3},
0)$.}} \vspace{.5cm} \centerline{\Large Figure 2}
\end{figure}
\hfill
\newpage
\newpage
\section*{\Large \sf References}

{\normalsize \sf

\noindent
[1] P.F. Harrison, D.H. Perkins, W.G. Scott, Phys. Lett. {\bf B530} (2002) 167.\\
X.G. He and A. Zee, Phys. Lett. {\bf B560} (2003) 87.\\

\noindent
[2] E. Ma and G. Rajasekaran, Phys. Rev. {\bf D 64} (2001) 113012 [arXiv: hep-ph/010629];\\
E. Ma, Mod. Phys. Lett. {\bf A17}(2002) 627 [arXiv: hep-ph/0203238];\\
G. Altarelli and F. Ferruglio, Nucl. Phys. {\bf B741}(2006) 215 [arXiv: hep-ph/0512103] and references therein.\\

\noindent
[3] Z.Z. Xing, Phys. Lett, {\bf B533} (2002) 85;\\
P.F. Harrison and W.G. Scott, Phys. Lett. {\bf B535}(2002) 163;\\

\noindent
[4] R. Friedberg and T.D. Lee, Ann. Phys. {\bf 324} (2009) 2196;\\
T. Araki and  C.Q. Geng, Phys. Rev. {\bf D79}, 077301 (2009) [arXiv:0901.4820v2].\\

\noindent
[5] C. Jarlskog, Phys. Rev. {\bf D35} (1987) 1685.\\

\noindent
[6] Super Kam Collaboration: Y. Ashie et al., Phys. Rev. Lett. {\bf 93} (2004)101801
[arXiv:0808.2016];\\
J. Hosaka et al., Phys. Rev. {\bf D 74} (2006) 032002;\\
K. Abe et al. Phys. Rev. Lett. {\bf 97} (2006) 171801.\\

\noindent
[7]Chooz collaboration: M. Apollonio et al., Eur. Phys. J. {\bf C27} (2003)331 [arXiv:hep-ex/0301017].\\

\noindent
[8] KamLAND collaboration: T. Araki et al., Phys. Rev. Lett., {\bf 94} (2005)081801 [arXiv:hep-ex/0406035].\\

\noindent
[9] MINOS collaboration: P. Adamson et al., Phys. Rev. Lett. {\bf 103} (2009) 261802 [arXiv:0909.4996].\\

\noindent
[10] G.L. Fogli et al., Phys. Rev. {\bf D78} (2008) 033010;\\
T. Schwetz et al., New J. Phys. {\bf 10} (2008) 113011;\\
Z.Z. Xing, Int. J. Mod. Phys. {\bf A23} (2008) 4355.\\

\noindent
[11] SAGE collaboration: J. N. Abdurashitov et al., Phys. Rev. {\bf C80} (2009)015907
[arXiv:0901.2200].\\

\noindent
[12] SNO collaboration: B. Aharmin et al., Phys. Rev. Lett. {\bf 101} (2008) 111501 and references therein.\\

\noindent
[13] Super Kam collaboration: J. Hosaka et al., Phys. Rev. {\bf D73} (2006) 112001 [arXiv:hep-ex/0508053].\\

\noindent
[14] S.F. Ge, H.J. He and F.R. Yin,  JCAP {\bf 05} (2010) 017 [arXiv:1001.0940].\\

\noindent
[15] Double-Chooz collaboration: C. Palomares [arXiv:0911.3227];\\
F. Ardellier et al., [arXiv:hep-ex/0405032];\\
F. Ardellier et al., [arXiv:hep-ex/0606025].\\

\noindent
[16] Daya Bay collaboration: X. Gue et al. [arXiv:hep-ex/070120];\\
S. Chen, J. Phys.:Conference Series {\bf 120} (2008)052024;\\
C. White, J. Phys.:Conference Series {\bf 136} (2008)022012.\\

\noindent
[17] J Parc collaboration: A.K. Ichikawa, Lect. Notes Phys. {\bf 781} (2009) 17.\\

\noindent
[18] Nova collaboration: D.S. Ayres et al. [arXiv:hep-ex/0503053].\\

}
\end{document}